\documentstyle[twoside,fleqn,espcrc2,epsf]{article}

\begin{document}

\title{The Deuteron Confronts Big Bang Nucleosynthesis}
\author{George M. Fuller and Christian Y. Cardall
\address{Department of Physics, University of California, San Diego,
La Jolla, California 92093-0319}}
\begin{abstract}
Recent determinations of the deuterium abundance, $^2$H/H, in high  
redshift Lyman limit hydrogen clouds challenge the usual picture of  
primordial nucleosynthesis based on \lq\lq concordance\rq\rq\ of the  
calculated light element ($^2$H, $^3$He, $^4$He, $^7$Li)  
nucleosynthesis yields with the observationally-inferred abundances  
of these species. Concordance implies that all light element yields  
can be made to agree with the observationally-inferred abundances  
(within errors) for single global specifications of the  
baryon-to-photon ratio, $\eta$; lepton number; neutron lifetime; and  
expansion rate (or equivalently, effective number of light neutrino  
degrees of freedom $N_{\nu} $). Though one group studying Lyman limit  
systems obtains a high value of $^2$H/H ($\sim 2\times {10}^{-4}$),  
another group finds consistently low values ($\sim 2\times  
{10}^{-5}$). In the former case, concordance for $N_{\nu} =3$ is  
readily attained for the current observationally-inferred abundances  
of $^4$He and $^7$Li. But if the latter case represents the  
primordial deuterium abundance, then concordance for {\it any}  
$N_{\nu}$ is impossible unless the primordial value of $^7$Li/H  
is considerably larger than the abundance of lithium as measured in  
old, hot Pop II halo stars. Furthermore, concordance with
$N_{\nu}=3$ is possible for low $^2$H/H 
only if either (1) the primordial $^4$He  
abundance has been significantly underestimated, or (2) new neutrino  
sector physics is invoked. We argue that systematic underestimation
of both the $^7$Li and $^4$He primordial abundances is the  
likely resolution of this problem, a conclusion which is strengthened  
by new results on $^4$He.            

\end{abstract}

\maketitle

\section{OVERVIEW}


In this paper we assess primordial nucleosynthesis
in light of the new Lyman limit system-derived deuterium abundance  
data \cite{rugers96,hogan95b,songaila94,burles96,tytler96}. 
Based on our analysis we adopt the  
view that currently we do not have a reliable handle on the  
primordial abundances of $^3$He, $^4$He, and $^7$Li; whereas, the  
Keck telescope may be providing us with a direct measurement of the  
abundance of the very fragile deuteron in relatively  
chemically-unevolved primordial material.

This is a somewhat radical view, given that much of the past  
discussion on big bang nucleosynthesis (BBN) has been predicated on  
the idea of \lq\lq concordance\rq\rq\---where a single global  
specification of the parameters characterizing Nuclear Statistical  
Equilibrium (NSE) freeze-out nucleosynthesis can lead the calculated  
abundances to agree with their observationally-inferred primordial  
values within errors. In a computation of NSE freeze-out  
nucleosynthesis \cite{wagoner67}, it is necessary to specify: (1)  
either the entropy-per-baryon $s$ (cosmic average 
in units of Boltzmann's constant $s \approx  
2.63\times {10}^{8}\, {\Omega}_b^{-1} h^{-2}$; where ${\Omega}_b$ is  
the baryon closure fraction and $h$ is the Hubble parameter in units  
of $100\,{\rm km}\,{\rm s}^{-1}\,{\rm Mpc}^{-1}$), or the  
baryon-to-photon ratio $\eta$ (cosmic average $\eta \approx  
2.68\times {10}^{-8}\, {\Omega}_b h^2$); (2) the spatial distribution  
of either of these quantities, both on scales smaller than the  
particle horizon at the epoch of NSE freeze-out and on super-horizon  
scales; (3) the three net lepton-to-photon numbers; (4) the ratio of  
the axial vector to vector weak interaction coupling constants  
(derived from the neutron lifetime); and (5) the expansion rate  
through the epoch of freeze-out. The expansion rate is determined by  
the energy density in the horizon. In turn, in the expected  
radiation-dominated conditions of BBN, the energy density is usually  
parametrized as an effective number of light neutrino species  
$N_{\nu}$, representing all relativistic particle degrees of freedom  
beyond those contributed by photons, electrons and positrons.

The usual procedure has been to compute nucleosynthesis yields as a  
function of $\eta$, for specifications of $N_{\nu}$ and the neutron  
lifetime, {\it assuming} that the entropy is homogeneously  
distributed on all scales and {\it assuming} that all net lepton  
numbers are small (${\  
\lower-1.2pt\vbox{\hbox{\rlap{$<$}\lower5pt\vbox{\hbox{$\sim$}}}}\ }  
\eta$) and neutrino masses are small and there are no other  
relativistic degrees of freedom. These simple Occam's razor  
assumptions are then \lq\lq justified\rq\rq\ by finding a concordant  
$\eta$ where all the independently-determined primordial abundances  
line up with their values predicted in the calculations ({\it cf.}  
Refs. \cite{walker91,smith93b}). Such concordance-based  
justification  has been touted as being all the more impressive and  
secure, given that the predicted and observationally-determined  
abundances range over some ten orders of magnitude.


\section{THE \lq\lq CRISIS\rq\rq}

The claimed precisions in the determinations of the primordial  
abundances of $^2$H, $^3$He, $^4$He, and $^7$Li have increased to the  
point where, if taken at face value, they invalidate the simple  
picture of concordance outlined above \cite{hata95a}.
This is the recent so-called \lq\lq Crisis\rq\rq\ in BBN. In fact,  
there were even earlier hints at a potential problem with the  
standard picture of concordance 
\cite{fuller91,kernan94}.  
With their adopted abundances of the  
$^2$H and $^4$He, the authors of Ref. \cite{hata95a} 
find no concordance for $N_{\nu} =3$, and derive a best fit to  
concordance for $N_{\nu} \approx 2.1 \pm 0.3$ with $N_{\nu} =3$ ruled  
out at the $98.6\%$ C.L.. These authors have suggested that perhaps  
this discrepancy could be eliminated with the introduction of new  
neutrino physics, essentially relaxing the usual assumptions  
regarding the above-discussed parameters (3) and (5) in BBN.

This conclusion and interpretation has been disputed by Copi {\it et.  
al.} \cite{copi95b}, who take as a prior assumption that  
$N_{\nu} =3$ and then argue that $^4$He has been underestimated.  
Cardall and Fuller (\cite{cardall96}, see also \cite{hatacosm96}) 
have done a re-analysis of  
this problem in light of the discordant determinations of $^2$H in  
Ly-$\alpha$ clouds and with special attention to the dependence of  
the BBN yields on $N_{\nu}$ and to the $^7$Li non-concordance problem  
(which they conclude cannot be rectified with new neutrino physics).   
This work tends to support the Copi {\it et al.} assessment. 
As we shall see, new work  
on the observationally-inferred abundances of $^2$H and $^4$He  
provides further support for this view and offers a hint of a new  
concordance.

\section{DISCORDANT DEUTERIUM MEASUREMENTS}

The deuteron is the most fragile of all nuclei, with a binding energy  
of only $E_B \approx 2.225\,{\rm MeV}$. As a result, the $^2$H yield  
in BBN is exponentially sensitive to $\eta$, though only mildly  
sensitive to $N_{\nu}$ \cite{wagoner67}. However, the fragile nature  
of the deuteron makes it extremely vulnerable to even small amounts  
of stellar processing and this, in turn, calls into question claims  
that we have a reliable handle on the primordial deuterium abundance.  
Direct measurements using the Keck telescope of isotope-shifted  
hydrogen lines in high redshift Ly-limit systems along lines of sight  
to distant QSO's may completely circumvent these stellar processing  
issues \cite{songaila94}. This is because these systems have  
manifestly low metallicity and this argues against significant  
stellar processing-induced destruction of $^2$H \cite{jedamzik96,malaney96}.

At present, however, there is no consensus on the primordial value of  
$^2$H/H from this technique: the Seattle-Hawaii group
\cite{rugers96,hogan95b,songaila94} obtains a very high range for this  
quantity ($15 {\  
\lower-1.2pt\vbox{\hbox{\rlap{$<$}\lower5pt\vbox{\hbox{$\sim$}}}}\ }  
d_5 {\  
\lower-1.2pt\vbox{\hbox{\rlap{$<$}\lower5pt\vbox{\hbox{$\sim$}}}}\ }  
23$ from Ref. \cite{hogan95b}); whereas, the San Diego group  
\cite{burles96,tytler96} examining different clouds obtains a  
consistently much lower range ($1.7 {\  
\lower-1.2pt\vbox{\hbox{\rlap{$<$}\lower5pt\vbox{\hbox{$\sim$}}}}\ }  
d_5 {\  
\lower-1.2pt\vbox{\hbox{\rlap{$<$}\lower5pt\vbox{\hbox{$\sim$}}}}\ }  
3.5$, combined results from Refs. \cite{burles96} and \cite{tytler96}  
with $\pm 2\sigma$ statistical error and $\pm 1\sigma$ systematic  
error). Here $d_5 \equiv ^2$H/H $ \times {10}^5$. 

An analysis of these discordant ranges 
is performed in Ref. \cite{cardall96}.
In this analysis, we have adapted the Kawano BBN code (with  
neutron lifetime from Ref. \cite{montanet94}, reaction rates from  
Ref. \cite{smith93b}, and a small, nearly $\eta$-independent  
correction to the $^4$He yield, $+0.0031$, from the time step and  
weak rate corrections of Ref. \cite{kernan94}). The  BBN yields for $^2$H,  
$^4$He, and $^7$Li are plotted 
as functions of $N_{\nu}$ and $\eta$. Such plots  
give insight into the leverage which each light element species has  
on concordance. Since, in contrary fashion to $^2$H, the BBN $^4$He  
yield is relatively sensitive to $N_{\nu}$ and much less sensitive to  
$\eta$, the tension between the overlap of these two species on our  
plots provides the most strigent criterion for concordance. 
This analysis shows  
that  for the typically  adopted range of the  
$^4$He abundance \cite{olive95c} and the \lq\lq Spite Plateau\rq\rq\ 
$^7$Li abundance measured in old, hot Pop II  
halo stars \cite{spite82,molaro95,ryan96},
concordance is readily attained for the  
Seattle-Hawaii $^2$H for $N_{\nu} =3$. On the other hand, 
no concordance with  
these values of $^4$He and $^7$Li is possible for {\it any} $N_{\nu}$  
with the San Diego deuterium determination.

Lack of concordance with the  \lq\lq Spite Plateau\rq\rq\ $^7$Li abundance 
may not be a serious problem: this  determination of primordial $^7$Li
is fraught with potential sytematic uncertainty, as  
$^7$Li is destroyed readily by $^7{\rm Li}({\rm p}, \alpha )\alpha$  
at temperatures as low as $T\sim 0.1\,{\rm keV}$. Therefore,  
rotation-induced mixing and turbulent diffusion could have destroyed  
most of the original $^7$Li on the surfaces of the old halo stars  
\cite{pins92,chaboyer94,deliyannis95}. However, the issue is complicated by  
the claimed observation of the even more fragile species $^6$Li in  
some of these objects\cite{hobbs94,smith93}. If mixing-induced  
destruction ({\it depletion}) of $^7$Li has indeed occurred, then the  
$^6$Li could have been produced {\it in  
situ} \cite{deliyannis95b}; alternatively, stellar wind-driven 
mass loss could deplete $^7$Li  
while leaving some $^6$Li present \cite{vauclair95}. 

We here present plots showing the concordance situation when
allowance is made for some $^7$Li depletion. 
In Figure 1 we show the $\eta$-$N_{\nu}$ parameter space  
corresponding to the Seattle-Hawaii deuterium range (dotted lines), 
along with the  
ranges for $^4$He ($0.223 \le {\rm Y_p} \le 0.245$,
solid lines) and $^7$Li  
($0.7\le l_{10} \le 3.8$, dashed lines) taken from Ref.   
\cite{olive95c}. Here $\rm Y_p$ is the primordial mass fraction of
$^4$He, and $l_{10} \equiv {^7{\rm Li}}/{\rm H} \times {10}^{10}$.  
This range of $l_{10}$ reflects $\pm 2\sigma$ statistical errors and $\pm  
1\sigma$ sytematic errors, and allows for a factor of $2$ depletion  
(contained in the sytematic error) on the Spite Plateau.
Concordance of all of these species is apparent for  
$N_{\nu} =3$ in Figure 1. 
However, if we adopt the San Diego deuterium range as  
primordial, with the same ranges adopted for the other species  
(Figure 2), then the \lq\lq crisis\rq\rq\ is evident, as there is no  
statistically significant overlap for $N_{\nu} =3$.

\begin{figure}
\epsfysize=7cm \epsfbox{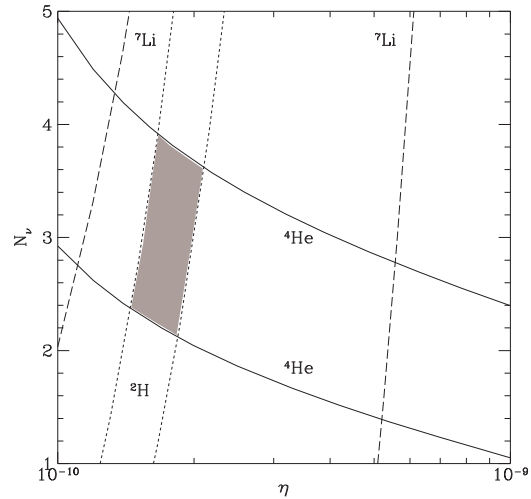}
\caption{Concordance plot for the Seattle-Hawaii $^2$H 
	determinations ($15 \le d_5 \le 23$, dotted lines) 
	and a typical $^4$He range ($0.223 \le {\rm Y_p} \le 0.245$,
	solid lines).}
\end{figure}

\begin{figure}
\epsfysize=7cm \epsfbox{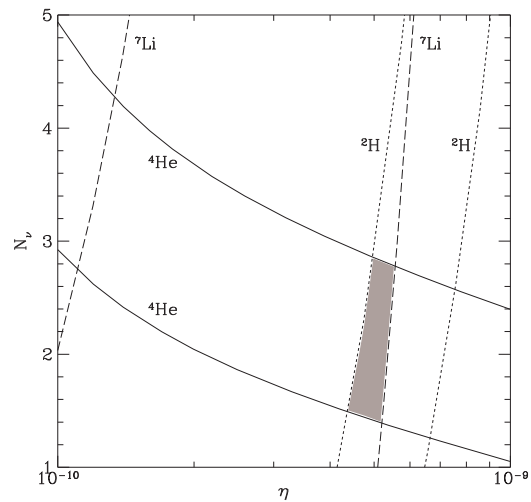}
\caption{Concordance plot for the San Diego $^2$H 
	determinations ($1.7 \le d_5 \le 3.5$, dotted lines) 
	and a typical $^4$He range ($0.223 \le {\rm Y_p} \le 0.245$,
	solid lines).}
\end{figure}

The quality of the San Diego group's data and the  
sophistication of their self-consistent analysis cannot be dismissed.  
Furthermore, there are a number of astrophysical problems which are  
partially or completely ameliorated if the San Diego group's  
deuterium range is adopted as primordial. This range for $^2$H/H  
would correspond to a range in baryonic closure fraction $0.016 {\  
\lower-1.2pt\vbox{\hbox{\rlap{$<$}\lower5pt\vbox{\hbox{$\sim$}}}}\ }  
{\Omega}_b h^2 {\  
\lower-1.2pt\vbox{\hbox{\rlap{$<$}\lower5pt\vbox{\hbox{$\sim$}}}}\ }  
0.03$. In contrast, adoption of the Seattle-Hawaii $^2$H/H as  
primordial would imply the much smaller closure fraction  $0.0056 {\  
\lower-1.2pt\vbox{\hbox{\rlap{$<$}\lower5pt\vbox{\hbox{$\sim$}}}}\ }  
{\Omega}_b h^2 {\  
\lower-1.2pt\vbox{\hbox{\rlap{$<$}\lower5pt\vbox{\hbox{$\sim$}}}}\ }  
0.0075$. There are hints from the x-ray galaxy cluster problem  
\cite{white93,white95} that the higher range for ${\Omega}_b  
h^2$ is to be preferred---{\it e.g.} the Coma cluster apparently has  
a fractional baryonic mass $f_b \approx 0.009 + 0.05 h^{-3/2}$ or  
$f_b \approx 0.15$ for $h=0.5$. Likewise, the MACHO project  
gravitational microlensing results \cite{alcock95,griest96}  
suggest that most or all of the galactic dark halo mass  
(corresponding to ${\Omega} \approx 0.02$ to $0.07$) is composed of  
objects with masses in the range $\approx 0.2\,{\rm M}_{\odot}$ to  
$\approx 0.6\,{\rm M}_{\odot}$. It is clear that the Seattle-Hawaii  
deuterium range implies a baryonic closure fraction which is  
difficult to reconcile with these MACHO results if the lensing  
objects had baryonic progenitors. Finally, observation of the cosmic  
background radiation Doppler peaks may require a high range of  
${\Omega}_b h^2$ \cite{silk96}.

An alternative to the adoption of the San Diego deuterium as  
primordial would be to invoke super-horizon scale entropy  
fluctuations at the epoch of BBN ({\it i.e.}, relax the homogeneity  
assumption of BBN parameter 2) \cite{jedamzik95,copi495,gnedin95}. 
Though such a scheme could give a comfortably high  
${\Omega}_b h^2\approx 0.05$ \cite{jedamzik95}, it would require a  
fair degree of fine tuning of the fluctuation spectrum and would {\it  
demand} that the cosmic average primordial deuterium abundance be  
high $^2{\rm H}/{\rm H} \sim {10}^{-4}$ \cite{jedamzik95}. Present  
statistics of Lyman limit sytems and other uncertainties would have  
to improve significantly to establish such intrinsic inhomogeneity  
\cite{jedamzik96}, so it seems reasonable to discount this scheme at  
present and see if a concordance can be found with adoption of the  
San Diego deuterium as a homogeneous primordial value.

\begin{figure}
\epsfysize=7cm \epsfbox{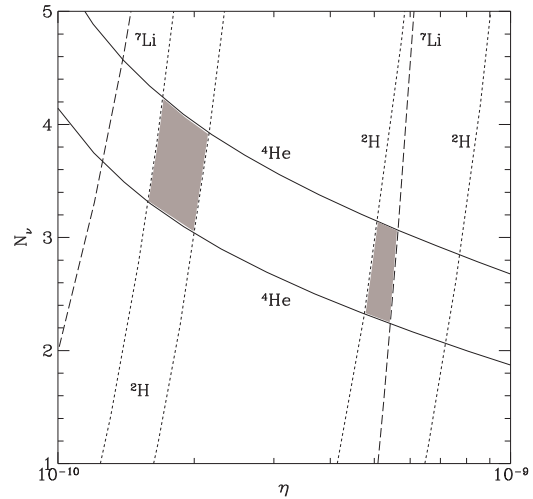}
\caption{Concordance plot showing both the Seattle-Hawaii
	and San Diego $^2$H ranges (dotted lines) and a new determination
	of the primordial $^4$He abundance ($0.237 \le {\rm Y_p} \le  
	0.249$, solid lines).}
\end{figure}

\section{NEW $^4$HE: A NEW CONCORDANCE}

A recent reinvestigation (with new data) of the linear regression  
method for estimating the primordial $^4$He abundance has called into  
question the systematic uncertainties assigned to ${\rm Y_p}$  
\cite{izotov96}. In fact, this study derives ${\rm Y_p} \approx 0.243  
\pm 0.003$, where the $1\sigma$ error is statistical. It is clear  
that the central value of this result is well above twice the  
systematic error assigned by Refs. \cite{hata95a,olive95c},  
confirming the suspicion \cite{cardall96,sasselov95} that  
${\rm Y_p}$ is not known well enough to draw sweeping conclusions  
regarding a lack of concordance. 

In Figure 3 we show both the San  
Diego and Seattle-Hawaii $^2$H ranges (dotted lines), along with the  
same $^7$Li range range (dashed lines) employed in Figure 1, but now  
with a band (solid lines) for $^4$He ($0.237 \le {\rm Y_p} \le  
0.249$, reflecting $\pm 2\sigma$ statistical errors and no systematic  
uncertainty) meant to be representative of the Ref. \cite{izotov96}  
results. It is evident from this figure that there is now no  
statistically significant concordance between $^4$He and the  
Seattle-Hawaii $^2$H for $N_{\nu} =3$, while there is now a hint of  
concordance for the San Diego $^2$H range for $N_{\nu} =3$. The new  
concordance engendered by the San Diego deuterium would be even  
better if allowance for systematic error were to be made in the  
$^4$He range. Such a new concordance would probably still require  
significant depletion of $^7$Li in old, hot Pop II halo stars  
\cite{cardall96}, though the classic constraints on neutrino physics  
from BBN \cite{walker91,copi95a,enqvist92,shi93,kang92} 
would survive intact and could be  
strengthened \cite{cardall96,cardall396}.
       
Supported by grants NSF PHY95-03384 and NASA NAG5-3062
at UCSD.

\end{document}